# A study on Modeling of Dependency between Configuration Parameters and Overall Energy Consumption in Wireless Sensor Network (WSN)

Najmeh Kamyabpour, Doan B.Hoang

iNext center for innovation in IT services and applications,
University of Technology, Sydney
{najmeh, dhhoang}@it.uts.edu.au

**Abstract.** In this paper, we study a new approach to model the overall Energy Consumption (EC) in Wireless Sensor Networks (WSN). First, we extract parameters involving in the EC of WSNs. The dependency between configuration parameters and the average residual energy of a specific application is then investigated. Our approach has three key steps: profiling, parameter reduction, and modeling. In profiling, a sensor network simulator is re-run 800 times with different values of the configuration parameters in order to profile the average residual energy in nodes. In the parameter reduction, three statistical analyses (p-value, linear and non-linear correlation) are applied to the outcome of profiled experiments in order to separate the effective parameters on WSN residual energy. Finally, linear regression is used to model the relation between the chosen effective parameters and the residual energy. The evaluation based on running the simulator for another 200 times with different values of the effective parameters shows that the model can predict the residual energy of nodes in WSN with average error of less than 13%.

## 1 Introduction

Recently, wireless sensors networks (WSNs) have attracted a huge amount of attention in many fields including environment, health, disaster alert, car, building, and mining industries. They also have a great potential to revolutionize many aspects of our lives. The networking of sensors is based on the fact that sensors are most useful when they are deployed in large numbers, especially for collecting environmental map of a geographical area such as a complete building, an agriculture field, or a rain forest. Once the sensors are deployed in a sensor application, they may no longer be accessible for further physical manipulation such as fixing faulty components or changing batteries. Therefore, it becomes important to manage the EC in order to increase the life time and consequently to maximize the performance of the sensor application. Clearly, to minimize the EC of wireless sensor networks, many inter-related factors must be considered. For example, the behavioral of the EC of an individual sensor is often forced by the target application. Also to deliver its data, the sensor has to rely on its neighbors to transmit its data to the destination. Therefore, the way sensors interconnect with together plays a critical role in determining the overall EC of the network. In addition to interconnection between sensors, sensing mechanisms, transmission mechanisms, networking protocols, topology, and routing all also play crucial parts in the overall energy consumption. These parts are often internally related in a complex network which makes the network difficult to analyze or optimize.

Efforts in minimizing the EC have increased over the last few years, however, they mostly focused on some specific and separate components of energy dissipation in WSNs such as MAC protocols [1], [2], routing [3], topology management[4] and data aggregation[5]. These components are, however, highly integrated within a WSN but their interplay cannot be taken into account as each component is treated independently without regard for other components. Minimizing the EC of one component, e.g. MAC protocols, may increase the energy requirements of other components like routing. Therefore minimizing energy in one component may not guarantee the minimization of the overall EC of the entire network. Also, a WSN can be seen as a system with several configuration parameters. The same as other systems [6], extracting the effective configuration parameters and tuning them properly have direct influence on estimating the EC behavioral of systems.

In this paper, we propose a method to extract most of the configuration parameters involved in influencing the EC followed by distinguishing the most influential parameters in term of their correlation. Then, a multiple linear regression is employed to model the relation between the most influential configuration parameters and the energy.

The rest of the paper is structured as follows. We first some related efforts in minimizing individual components of WSNs are summarized, in section 2, followed by a discussion over problem statement in section 3. An explanation on multiple linear regression modeling will be held in section 4. Experimental evaluation will be in section 5. Finally, we summarize our work and outline future research directions in section 6.

## 2. Related Work

Network architectures such as OSI and Internet are basically functional models organized as layers where the below layer provides services to the above layer (e.g. the application layer provides services to the end users). Network is often evaluated in terms of its quality of service parameters such as delay, throughput, jitter, availability, reliability and even security. However, when it comes to the EC, one often encounters difficulty in evaluation and optimization of the network as a comprehensive model that takes the EC into account hardly exists. Generally, researchers focus on the traditional network architecture and try to minimize specific component of a single layer with the hope that the overall EC of the network is reduced without regard for other components or layers. This is almost not an ideal situation where one does not know how a single component fits within the overall energy picture of an entire wireless sensor network. Most current energy minimization energy models focus on send and receive data[7], while other parameters influencing them are neglected. In [8]and[9] the power consumption model focused on the cost of sending and receiving data and deduced the upper limit of the energy efficiency of single hop distance. This approach considers an intermediate node between source and destination so that the retransmission will save the energy. Other approaches evaluate the energy efficiency of wireless sensor network by using the power consumption model mentioned in[8]and[9].

Since Wireless networks has different specification and challenges, the traditional network architecture cannot satisfy them. Cross layer idea is being created to provide flexible network architecture for Wireless Networks. The key idea in cross-layer design is to allow enhanced information sharing and dependence between the different layers of the protocol stack[10], [11], [12]. However it is argued that by doing so, performance gains can be obtained in wireless networks since the resulting protocols are more suited to be employed on wireless networks as compared to protocols designed in the strictly layered approach. Broad examples of cross-layer design include, say, design of two or more layers jointly, or passing of parameters between layers during run-time etc but there is no criteria to determine which layers should be combine to give the best result for the overall EC[13].

In figure 1, we show the overall power usage in term of all possible energy consuming constituents and their configuration parameters[14-15] ranging from hardware parameters (like Dynamic-voltage frequency scaling [16][23]) to higher level parameters. Configuration parameters may directly affect on number of defined tasks for constituents. Moreover, it is too probable changing a parameter value decreases tasks in one constituent while increasing tasks in other constituents. Therefore, it becomes important to find the dependency between the configuration parameters and the power consumption. This dependency can help the researchers to focus on energy-efficient approaches in the whole system instead of the optimizing energy in a specific layer. In addition we define sensor tasks in constituent basis format and consider the value of constituent's parameters as effective factor in the EC of a sensor. This view point introduces a modular solution to analysis the overall EC of Wireless sensors Networks. In this paper, we utilized non-linear correlation to analysis the possible dependencies between effective parameters and the energy. Finally we purpose a power consumption model for a deployed sensor in a WSN.

## 3. Problem Statement

Among several challenging problems in wireless sensor networks, the energy is one of the biggest as it has a great dependency on the life time of a network. If a sensor in WSN can manage efficiently its battery resource so that it can work for longer time. Among all sensors in a network, the one which engages in several activities such as routing has less life time than others. This is why energy-efficient routing algorithms are one of the hottest problems in WSN. There are a huge number of research works in literature on minimizing the EC on different network layers and levels. To best of our knowledge, there is not an overall EC model involving effective network parameters. Moreover there is not any approach to optimize the EC of the whole network. Existing approaches only focus on a specific layer. Such a model is necessary as it is possible a component influences more than one layer of the whole network. This means if a component with a specific approach minimize the energy in one layer, it may increases the energy of other layers which may result in increasing the overall network energy usage.

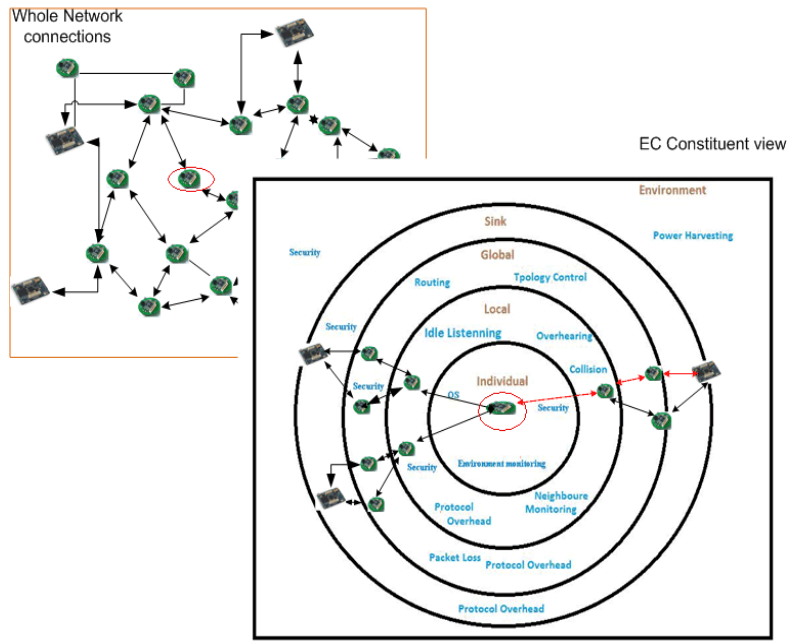

**Figure 1. Sensor centric view of a Wireless Sensor Network**

In response to this problem, we should first answer how energy is consumed in WSNs. We assume a modular view including all effective components on energy (Figure 1). This view point is specifically useful when one wants to optimize the whole energy in term of the tasks defined for sensors in the network. This model classifies the overall EC into five task-based constituents as shown in figure 1: individual, local, global, environment, and sink. Parameters associated with constituents can then be extracted. Individual constituent defines all the essential and basic operations or tasks for the sensor to just exist i.e. monitoring environment events as a key task of a sensor, executing OS and providing security in OS level while local constituent deals with initiating and maintaining all communications between a node's immediate neighbors i.e. monitoring neighbors and providing security for communication in neighbor local level. In addition, it may include power usage of overhearing, idle listening and collision if they happen. In global constituent, the concern is with the maintenance of the whole network, the selection of a suitable topology and an energy efficient routing strategy based on the application's objective.

This may include energy wastage from packet retransmissions due to congestion and packet errors. The global constituent is defined as a function of the EC for topology management, packet routing, packet loss, and protocol overheads. Sink constituent assumes the roles of manager, controller or leaders in WSNs. The sink tasks include to direct, balance and minimize the EC of the whole network and to collect the generated data by the network's nodes. In contrast to other constituents which spend energy, environment constituent consists on deploying the harvesting operation in the case that nodes have capability of extracting energy from environment.

The sensor battery lifetime depends on how many tasks involve in Individual, local, global, environment and sink constituents. To execute a task the sensor needs to exchange a number of packets. Number of packets for Constituent's tasks may change based on Constituents Input parameters. Tables 1-5 (appendix) show the extracted parameters involved in WSN based on the constituents. Although this model gives a complete view about the parameters, modeling of such a system when the number of parameters is huge (34 parameters) becomes very difficult or maybe impossible. In this paper, we clearly follow these steps:

- Use statistical dependency analysis to extract the most effective parameters on residual energy. The method we are using is based on three mathematical methods: p-value, linear and non-linear correlation. P-value is generally used to find the effective parameters which correlation is utilized to also find the sign of the dependency. As some parameters may have a nonlinear relation with energy, non-linear correlation is also used to improve the dependency analysis.
- After reducing the parameters, a multiple linear regression is used to model the dependency between effective parameters and the residual energy.

## 4. Energy-related Dependency Analysis and Modeling

The analytical part of the proposed method in this paper includes two mathematical approaches: (1) parameter reduction by analyzing dependency between all parameters and average residual energy in network and consequently reduce the number of parameters by keeping the parameters with correlation value more than a pre-defined threshold. (2) Use linear regression to model the relation. Obviously, the model with parameter reduction should have less accuracy but high simplicity than a model without parameter reduction.

### 4.1 Parameter reduction

P-value and correlation analysis are the most common ways in statistics to analysis the dependency between two random variables. As there may be nonlinear relation between some of the parameters and the output (here average residual energy), non-linear correlation analysis is also used with p-value and linear correlation analysis. As the highest values of correlation and the sign of the correlation are more important than the magnitude of correlation is analysis, linear correlation is utilized in our paper.

Furthermore, P-value analysis is also utilized to verify the result of correlation. In the sense that if both analyses show the same highest correlated variables, so the result is approved. In another word, p-value is specifically used to obtain the most statistically significant variables influencing the output variable and linear correlation is used for both influential variables and the variables direction effect.

a. **P-value**

Making a decision about statistical significance is related to the practice of hypothesis testing. In general, the idea is to state a null hypothesis (i.e. that there is no effect) and then to see if the gathered data allows you to reject the hypothesis. In statistics, a null hypothesis ($H_0$) is tested by gathering data and then measuring how probable is the occurrence of the data, assuming the null hypothesis is true. If the data are very improbable (usually defined as observed less than 5% of the time), then the experimenter concludes that the null hypothesis is false. If the data do not contradict the null hypothesis, then no conclusion is made. In this case, the null hypothesis could be true or false; the data give insufficient evidence to make any conclusion.

In statistics, p-value is a numerical measure of the statistical significance of a hypothesis test. If the probability of affecting the output by a specific input variable from a set of experiments is very low, the null hypothesis can be rejected otherwise, the dependency of the two is statistically significant. In our study, the null hypothesis is that to see if a configuration parameter of WSN has sufficient influence on the average residual energy. In another word, it tells how likely it is that we could have gotten our sample data even if the null hypothesis of correlation between a specific configuration parameter ($p_k$) and the average residual energy ($E$) is true. By convention, if the p-value is less than 0.05 (or $prob(reject\ H_0|H_0\ is\ valid) < 0.05$), we conclude that the null hypothesis can be rejected (i.e., $p_k$ and $E$ are highly correlated). In other words, when prob < 0.05 it is said that the dependency between $p_k$ and $E$ are statistically significant.

b. **Linear Correlation analysis**

Although p-value analysis is good enough to pick the most effective configuration parameters, it does not give any information about how the parameter and residual energy are dependent. Precisely, p-value does not give any information about both value and sign of the correlation. Therefore, correlation analysis is utilized accompany with p-value to get more information.

By definition, correlation is a measure of dependency between two variable sequences on a scale from -1 to 1. If $\boldsymbol{p_1} = \left(p_1^{(1)}, p_1^{(2)}, \ldots, p_1^{(M)}\right)$ and $\boldsymbol{E} = \left(E^{(1)}, E^{(2)}, \ldots, E^{(M)}\right)$ are constituents' parameters and the residual energy, respectively, the normalized cross-correlation function between these two series can be expressed as follows:

$$Corr_{norm}(\boldsymbol{p_1}, \boldsymbol{E}) = \frac{\sum_{i=1}^{M}\left(\left(p_1^{(i)} - \mu_{p_1}\right)\left(E^{(i)} - \mu_E\right)\right)}{\sqrt{\sum_{i=1}^{N}\left(p_1^{(i)} - \mu_{p_1}\right)^2 \sum_{i=1}^{N}\left(E^{(i)} - \mu_E\right)^2}} \quad (3)$$

Where $\mu_{p_1}$ and $\mu_E$ are the mean of $\vec{p_1}$ and $\vec{E}$ series, respectively. A correlation value of 0 indicates a random or independent relationship between the parameter and average residual energy, and the correlation values of 1 and -1 denote positive and negative respectively perfect linear associations between them.

**c. Non-linear Correlation analysis**

As the nature of the dependency between a WSN parameter and the output is unknown, so non-linearity between them is highly probable. As the ordinary correlation is not sufficient[17] for nonlinear cases, a nonlinear correlation analysis has been utilized based on the study in[18] and[19]. Here a brief description in[20] is described as a typical solution. The normalized higher order :

$$Corr_{norm}^2(\mathbf{p_1}, E)$$
$$= \frac{\sum_{i=1}^{M}\left(\left(\left(p_1^{(i)}\right)^2 - \frac{1}{M}\sum_{k=1}^{M}\left(p_1^{(k)}\right)^2\right)\left(\left(E^{(i)}\right)^2 - \frac{1}{M}\sum_{k=1}^{M}\left(E^{(k)}\right)^2\right)\right)}{\sqrt{\sum_{i=1}^{M}\left(\left(\left(p_1^{(i)}\right)^2 - \frac{1}{M}\sum_{k=1}^{M}\left(p_1^{(k)}\right)^2\right)\right)^2 \sum_{i=1}^{M}\left(\left(\left(p_1^{(i)}\right)^2 - \frac{1}{M}\sum_{k=1}^{M}\left(E^{(k)}\right)^2\right)\right)^2}} \quad (3)$$

It should be noticed that $Corr_{norm}^2(\mathbf{p_1}, E) = 1$ represents perfect dependency and $Corr_{norm}^2(\mathbf{p_1}, E) = 0$ represents complete independence between two variables. Obviously for M=1, the linear correlation is calculated.

**4.2 Model Generation**

This section describes how to model the relation between configuration parameters and average residual energy in WSN. The problem of modeling a WSN based on linear regression involves choosing the suitable set of of the modeling such that the model's response approximates well the real system response.

Consider the linear algebraic equations for $M$ experiments of an application for $N$ effective configuration parameters $(M \gg N)$ [21-22]:

$$\begin{cases} E_{t_1,\Delta t}^{(1)} = \alpha_0 + \alpha_1 p_1^{(1)} + \cdots + \alpha_N p_N^{(1)} \\ E_{t_1,\Delta t}^{(2)} = \alpha_0 + \alpha_1 p_1^{(2)} + \cdots + \alpha_N p_N^{(2)} \\ \quad\vdots \\ E_{t_1,\Delta t}^{(M)} = \alpha_0 + \alpha_1 p_1^{(M)} + \cdots + \alpha_N p_N^{(M)} \end{cases} \quad (1)$$

Where $E_{t_1,\Delta t}^{(k)}$ is the value of average residual energy in WSN in $k^{th}$ experiment on simulator and $\left(p_1^{(k)}, p_2^{(k)}, \ldots, p_N^{(k)}\right)$ are the values of $N$ chosen effective parameters for the same experiment, respectively. Eqn.1 can be rewritten in matrix format as,

$$\underbrace{\begin{bmatrix} E_{t_1,\Delta t}^{(1)} \\ E_{t_1,\Delta t}^{(2)} \\ \vdots \\ E_{t_1,\Delta t}^{(M)} \end{bmatrix}}_{E_{t_1,\Delta t}} = \underbrace{\begin{bmatrix} 1 & p_1^{(1)} & p_2^{(1)} & \cdots & p_N^{(1)} \\ 1 & p_1^{(2)} & p_2^{(2)} & \cdots & p_N^{(2)} \\ & & \vdots & & \\ 1 & p_1^{(M)} & p_2^{(M)} & \cdots & p_N^{(M)} \end{bmatrix}}_{P} \underbrace{\begin{bmatrix} \alpha_0 \\ \alpha_1 \\ \vdots \\ \alpha_N \end{bmatrix}}_{A}$$

Using the above formulation, the approximation problem transformed in to the problem of estimation of the values of $\widehat{\alpha_0}, \widehat{\alpha_1}, \ldots, \widehat{\alpha_N}$ to optimize a cost function between the approximation and real values of average residual energy. An approximated residual energy $\left(\widehat{E_{t_1,\Delta t}^{(k)}}\right)$ of the network for the $j^{th}$ experiment is then predicted as

$$\widehat{E_{t_1,\Delta t}^{(k)}} = \widehat{\alpha_0} + \widehat{\alpha_1} p_1^{(j)}, \ldots, \widehat{\alpha_N} p_N^{(j)} \qquad (2)$$

There are many of well-known mathematical methods for calculating the variables $\widehat{\alpha_0}, \widehat{\alpha_1}, \ldots, \widehat{\alpha_N}$. One of these methods used widely in computer science and finance is the Least Square Regression which calculates the parameters in Eqn.2 by minimizing the least square error as follows:

$$LSE = \sqrt{\sum_{i=1}^{M} \left(\widehat{E_{t_1,\Delta t}^{(i)}} - E_{t_1,\Delta t}^{(i)}\right)^2}$$

The set of coefficients $\widehat{\alpha_0}, \widehat{\alpha_1}, \ldots, \widehat{\alpha_N}$ is the model that describes the relationship between the average residual energy of the network and the configuration parameters. In other words, the approximate model between the average residual energy of the simulated WSN and the configuration parameters is:

$$E_{t_1,\Delta t} = \mathcal{F}(p_1, p_2, \ldots, p_N) \cong \widehat{\alpha_0} + \widehat{\alpha_1} p_1 + \cdots + \widehat{\alpha_N} p_N \qquad (3)$$

It can be mathematically proved that the least square error between real and approximated values is minimized when $A = (P^T P)^{-1} P^T E_{t_1,\Delta t}$. Once a model has been created, it can then be applied to a WSN experiment to estimate what the average residual energy in the network after $t$ seconds will be if the values of effective configuration parameters change.

## 5 Experimental Evaluation

### 5.1 Experimental Setting

We have simulated a wireless sensor application to track the energy usage of sensors. The application collects information about events that randomly appear in the screen. Sensors detect event in their covered area and create packet and send the packet to the nearest sink. Sinks are located as a group in a specific location. Sensors perform defined tasks in the application i.e. sensing, neighbor monitoring, relay data, routing, relay data if they have enough power otherwise they stop.

TABLE 6. Linear, non-linear and p-value analysis of constituents' effective parameters and Residual Energy

| Parameters | P-value | Linear Correlation | Non-Linear Correlation |
|---|---|---|---|
| *Transmission interval* | 3.7979e-005 | 0.2842 | 0.2474 |
| *Num hop* | 0.00051 | -0.2411 | -0.2247 |
| *Sensor interval* | 0.02397 | 0.1580 | 0.1280 |
| *Sense Radius* | 0.04933 | -0.1355 | -0.1178 |
| *Net density* | 0.11896 | -0.1095 | -0.0474 |
| *Transmission Radius* | 0.32401 | -0.0694 | -0.0693 |
| *Sink* | 0.42896 | -0.0557 | 0.0004 |
| *Neigh* | 0.44191 | -0.0541 | 0.0088 |

In this section the aim is to conduct the range of simulation experiments to model the EC with respect to the most effective parameters. First we shows which parameters are more significant and effective based on their p-value. We evaluated the EC during an interval time Δt. In particular we will focus on the individual, the local, and the global constituents. We focus at this stage extracting all possible application parameters and we randomly assign different values to them in different experiment. For the individual constituent, we have the sensor's sensing radius as it determines the coverage of the sensor field and sensor interval as the delay between two sense operations. For the local constituent, we select the transmission radius of a sensor as it concerns the number of sensors in covered area and number of neighbors that sensor collaborate with them and average distance between sensor and neighbors. Also we select transmission interval as the interval time between two following transmission operations. For the global constituent, we select number of hops, network density, number of sink scheme as it affects the data transport from sensors to sinks. We will investigate the influence of the individual, the local and the global constituents by measuring the overall EC.

### 5.2 Results

The effectiveness of configuration parameters and consequently constituents changes in different wireless sensor application. Based on p-value of parameters (Table 6), the Transmission interval, the Number of hops, the sensor interval, and the sense radius are the most effective parameters in our application because of their lowest p-values. The negative p-values show the reverse relationship between the parameter and the residual energy. For example if the number of hops increases the residual energy will decrease and consequently the overall EC will rise.

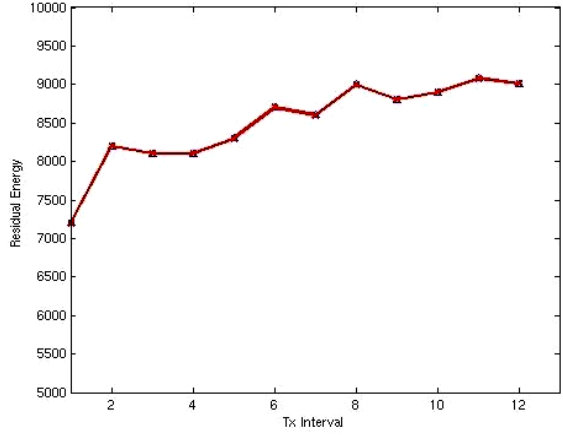

(a)

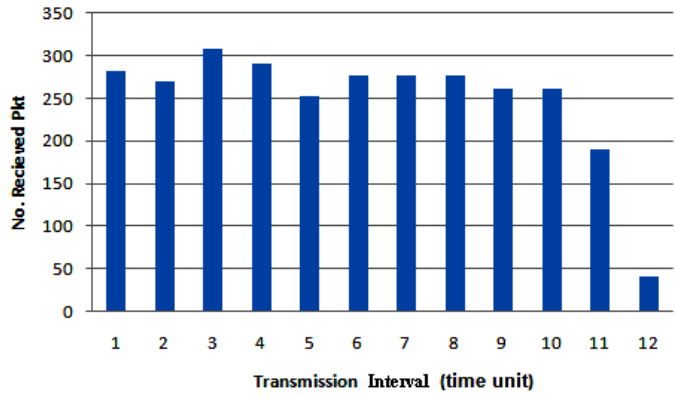

(b)

Figure 2. (a) Variations of average Residual Energy caused by number of hops (b) Network performance in term of number of received data packets in the sinks with different Transmission interval

The values of nonlinear and linear correlation neither observed strong linear nor nonlinear relationship (Table 6). Both values do not precisely prove the linearity or nonlinearity of the relationship between the parameters and the EC. We model the overall EC of a sensor using linear regression in term of the most effective parameters as follows:

$$E = \alpha_0 + \alpha_1 g_{Tx} + \alpha_2 h + \alpha_3 g_{sens} + \alpha_4 r_{sens}$$

$\alpha_1, \alpha_2, and\ \alpha_3$ are learned from many experiments with random values for the effective parameters. Figure 2a and 3a shows the relation between the network residual energy and the most effective parameters, transmission interval and number of hops.

In the application, sensors create a data packet when they detect an event in their area. The performance of the application is calculated based on number of detected events in the application. Therefore if $n$ events occur, in the good performance situation we will receive $n$ Data packet. We calculate $N$, as average number of events in Δt based on running many experiment. We consider variance, $v$, to measure the amount of variation of the number of events from the average number of events, $N$, in Δt. The variance determines boundary to have a good performance experiment in term of how far the number of detected events deviate from the average number of events. We consider the application performance as an important factor to study the influence of variation of parameters on the EC.

Transmission interval is the time between two consecutive transmission operations. If data packets are produced faster than the transmission interval the node put them in the buffer until it can get around to transmitting them. The maximum transmission interval is proportional to the node buffer size and the average event interval time. In our application, we assume nodes with $n$ packet buffer capacity. Therefore the

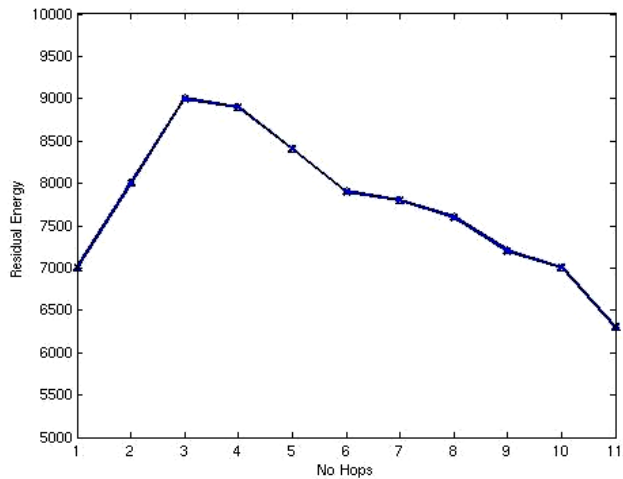

(a)

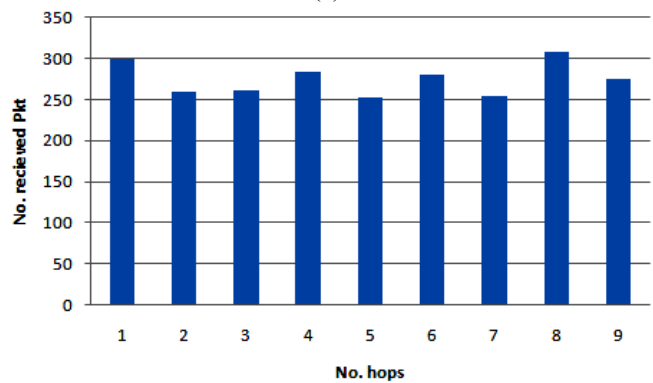

(b)

Figure 3. (a) Variations of average Residual Energy caused by Transmission interval (b) Network performance in term of number of received data packets in the sinks with different number of hops.

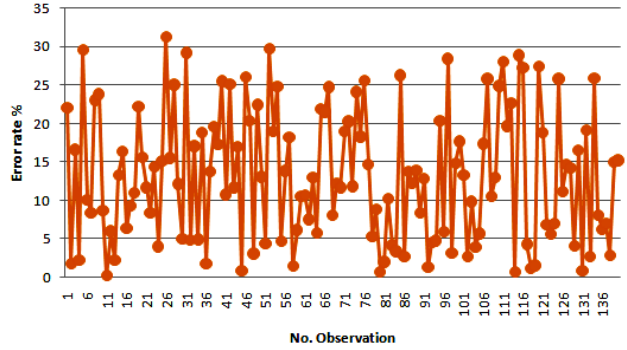

**Figure 4. The error range of the model**

transmission interval should satisfy following constrain to avoid packet drop:
$$T < g_{Tx} < nT$$

Where, $T$ is the average event interval time and n is buffer size. Also we have $nT \ll \Delta t$. As can be seen in figure 2a the average residual energy increases if transmission interval goes up. The direct relationship between residual energy and transmission interval also can be achieved from positive sign of the linear correlation in table 6. Since number of events and their position are random in the environment, we see fluctuation of the average residual energy in the experiments based on application performance the best value for the transmission interval can be defined in the boundary [2, 10] in term of the time unit. When the transmission interval value reach the upper, $nT$, nodes will send less packet in $\Delta t$ and packets dropped because of the buffer overflow and consequently as can be seen in figure 2b the application performance decreases. Therefore we can save the energy by assigning an optimum the transmission interval and with respect to the performance.

Furthermore, figure 3a shows if average number of hops goes up, the residual energy decreases due to dramatically increase in the number of control packets. Consequently nodes consume more energy to send data packets. However, we have significant increase of the EC when numbers of hops are very small e.g. none, one or two hops due to square relationship between the energy and the distance. The reverse relationship between the residual energy and the number of hops also can be resulted from negative sign of the linear correlation in table 6. As can be seen in the figure 3b the variations in the performance are negligible among various numbers of hops; however assigning a large number of hops can decrease the number of received packets in the sinks. Because finding the lowest number of hops that can satisfy the performance is the aim we do not determine the bound of the hops.

However the variation of the transmission interval and the number of hops plays a crucial rule in the EC, we should not ignore influence of other parameters because they all affect on energy in different ways or they may effect on either effective parameters or each other. For instance, a suitable value for the sensor radius based on the environment and the sensor capability can decrease redundancy in the network and consequently the EC for the redundant packet recognition and the transportation of redundant packets through the network will be saved. Furthermore, the sensor radius and the sensor interval as two other top important parameters can effect on the boundary of the transmission interval in term of the average number of data packets produced by the sensor in the experiment interval time Δt.

Figure 4 shows the prediction accuracy of our application by comparing the actual EC and its predicted value. We found that the average error between the observed EC and the predicted values is about 13%. The errors are expected from the model inaccuracy, the linearity assumption, and eliminating the influence of other parameters.

## 6  Conclusion

In this paper, we proposed a new approach to model the Overall EC with respect the most effective parameters. We defined energy consuming constituents and their effective parameters. Due to the high number of parameters, a reduction phase by using statistical p-value and correlation techniques has been used in order to analysis the dependency between the parameters and energy and therefore to keep the most effective parameters which expected to be more less than total number of parameters. After reduction, least square regression is applied to find the relation between the effective parameters and the network residual energy. Our evaluation on 1000 experiments shows that our approach proofs the importance of network configuration parameters on energy in the network. For the experiment result the problem of modeling between configuration parameters and the average residual energy of an event-based WSN has been studied.

# Appendix

TABLE 1. INDIVIDUAL EFFECTIVE PARAMETERS ON ENERGY CONSUMPTION

| | Individual Parameters | | |
|---|---|---|---|
| index | Parameter | Description | Boundary |
| 1 | $r_{sense}$ | Sensing radius points to the covered area of the sensor, this will have different meaning in different applications e.g. a temperature application and a radar application. | $r_{sense} > 0$ |
| 2 | $g_{sense}$ | Sensing delay | $g_{sense} \geq 0$ |
| 3 | $b_{sense}$ | Number of packet created by sensor itself that includes environment's data. | $b_{sense} \geq 0$ |
| 4 | $b_{store}$ | Numbers of packets are stored in the memory. | $b_{store} \geq 0$ |
| 5 | $b_{Os}$ | Number of Os instruction | $b_{OS} \geq 0$ |
| 6 | $b_{sec}$ | Security in Individual level | $b_{sec} \geq 0$ |

TABLE 2. LOCAL EFFECTIVE PARAMETERS ON ENERGY CONSUMPTION

| | Local Parameters | | |
|---|---|---|---|
| index | Parameter | Description | Boundary |
| 1 | n | Number of neighbors | $n \geq 1$ |
| 2 | $e_i(idle)$ | Idle power consumption | |
| 3 | $d_{ij}$ | Distance to the neighbor | $0 < d_{ij} \leq r_{Tx}$ |
| 4 | $b_{mon}$ | Packet overhead for monitoring depends on the application and its topology. | $b_{mon} \geq 0$ |
| 5 | $r_{Tx}$ | Transmission Radius | $r_{Tx} \geq 0$ |
| 6 | $b_{sec}$ | Local Security packet overhead depends on application. | $b_{sec} \geq 0$ |
| 7 | $b_{local}$ | Packet overhead to avoid collision problem policy. | $b_{local} \geq 0$ |
| 8 | $b_{reTx}$ | Number of retransmission packets depends on probability of collision and number of neighbors | $b_{reTx} \geq 0$ |

TABLE 3. GLOBAL EFFECTIVE PARAMETERS ON ENERGY CONSUMPTION

| | Global Parameters | | |
|---|---|---|---|
| index | Parameter | Description | Boundary |
| 1 | n | Number of neighbors | $n \geq 1$ |
| 2 | $g_{Tx}$ | Transmission interval | Application dependent |
| 3 | $net_{dens}$ | Network density | $2 \leq net_{dens}$ |
| 4 | $b_{ohear}$ | Overheard packets | $0 \leq b_{ohear}$ |
| 5 | $a_i$ | Number of sensors in covered area | $n_i \leq a_i$ |
| 6 | $d_{iA}$ | Distance between sensors and other sensors inside the covered area | $0 \leq d_{iA}$ |
| 7 | $b_{topo}$ | Packet overhead for topology | $b_{topo} \geq 0$ |
| 8 | $N(t)$ | Number of nodes in time t | $n_{snk}+2 \leq N(t)$ |
| 9 | $d_{iD}$ | Distance between source and destination. | $0 < d_{iD}$ |
| 10 | $h_{iD}$ | Number of Hops | $0 \leq h_{iD} < net_{dens}-1$ |
| 11 | $b_{rout}$ | Number of routing packets | $b_{rout} \geq 0$ |
| 12 | $b_{global}$ | Number of Packet to Avoid pktls problem | $b_{global} \geq 0$ |
| 13 | $d_i$ | Distance between node i to nearest sink | $d_i > 0$ |
| 14 | $b_{pktls}$ | Number of packet losses | $b_{pktls} \geq 0$ |
| 15 | Snk | Number of sinks | Snk>0 |
| 16 | $b_{sec}$ | PF security in global level | $b_{sec} \geq 0$ |

TABLE 4. ENVIRONMENT EFFECTIVE PARAMETERS ON ENERGY CONSUMPTION

| | Environment Parameters | | |
|---|---|---|---|
| index | Parameter | Description | Boundary |
| 1 | $H_i$ | Harvested energy (Wat) | $H_i \geq 0$ |
| 2 | $b_{ph}$ | Overhead produced due to harvesting power. | $b_{ph} \geq 0$ |

TABLE 5. SINK EFFECTIVE PARAMETERS ON ENERGY CONSUMPTION

| | Sink Parameters | | |
|---|---|---|---|
| index | Parameter | Description | Boundary |
| 1 | $b_{ohead}$ | Network management policy | $b_{ohead} \geq 0$ |
| 2 | $b_{sec}$ | PF security in sink level | $b_{sec} \geq 0$ |